\documentclass[12pt,a4paper,twoside]{article}
\usepackage{epsfig}

\newcommand{\be}{\begin{equation}}
\newcommand{\ee}{\end{equation}}
\newcommand{\beqs}{\begin{eqnarray}}
\newcommand{\eeqs}{\end{eqnarray}}
\newcommand{\beqsn}{\begin{eqnarray*}}
\newcommand{\eeqsn}{\end{eqnarray*}}
\newcommand{\da}{{\dot a}}

\newcommand{\p}{\mbox{$\Delta$}}
\newcommand{\pvv}{\mbox{$\Delta_{2v}$}}
\newcommand{\pmvv}{\mbox{$\Delta_{-2v}$}}
\newcommand{\pv}{\mbox{$\Delta_{v}$}}
\newcommand{\pmv}{\mbox{$\Delta_{-v}$}}

\newcommand{\di}{\mbox{$\partial_1$}}
\newcommand{\dii}{ \mbox{$\partial_2$}}
\newcommand{\bea}{\begin{eqnarray}}
\newcommand{\eea}{\end{eqnarray}}
\expandafter\ifx\csname mathbbm\endcsname\relax

\else

\fi
\begin{document}
\begin{titlepage}
\begin{flushleft}  
       \hfill                      {\tt hep-th/9807071}\\
       \hfill                      UUITP-3/98\\
       \hfill                       July 1998\\
\end{flushleft}
\vspace*{3mm}
\begin{center}
{\LARGE Graviton scattering in an M-orientifold background
\\}
\vspace*{12mm}
{\large 
Ulf H. Danielsson\footnote{E-mail: ulf@teorfys.uu.se} \\
\vspace*{5mm}
Mart\'{\i}n Kruczenski\footnote{E-mail: martink@teorfys.uu.se} \\
\vspace*{5mm}
P\"{a}r Stjernberg\footnote{E-mail: paer@teorfys.uu.se}\\
\vspace{5mm}

{\em Institutionen f\"{o}r teoretisk fysik \\
Box 803\\
S-751 08  Uppsala \\
Sweden \/}}
\vspace*{15mm}
\end{center}
%\maketitle

\begin{abstract}
In this paper we study the scattering of gravitons off a five orientifold in
eleven dimensions. We compare the supergravity result with a two loop
M(atrix) model calculation and find exact agreement. The supergravity
calculation involves nonlinear three graviton effects.
\end{abstract}

\end{titlepage}

\section{Introduction}
\label{sec:int}

The M(atrix)-model, \cite{bfss},
was originally introduced as a model for eleven
dimensional M-theory based on D-particle quantum mechanics, \cite{dfs,kp,dkps}.
To obtain the eleven dimensional theory it was important
to take the limit $N \rightarrow \infty$.
A stronger 
conjecture was however put forward in \cite{dlcq}, where it was proposed that
finite
$N$ makes sense as a lightlike compactification, the 
discrete light cone quantization (DLCQ). In 
\cite{seibergdlcq,sendlcq} strong 
arguments were given in favor of the DLCQ conjecture.
For reviews with extensive
references, see, for example \cite{banksreview,susskindreview,taylorreview,bilal}.

It is natural to test the conjecture by comparing supergravity with
perturbative
M(atrix) model calculations. If we want to compare with noncompact
$D=11$ supergravity we must make sure that the compact lightlike
direction is large enough.
According to \cite{svart1,svart2}, this requires a
large boost so that the physical dimensions of the system under study
are
much smaller than the compactification scale. This in turn implies
taking
$N$ very large. To be more explicit, consider a system of characteristic 
mass $M$ and characteristic length
$r$. If the system is at rest we have $P_- = P_+ =M$.
According to the argument above, we need $r << R$ (where $R$ is the
lightlike compactification radius) in order to have a reliable
description. This implies $Mr << MR = P_- R=N$. 
Given $r$, the largest mass that we need to consider is the mass of a
black hole of
radius $r$. In eleven dimensions we have $M \sim r^8$ and therefore we
conclude that
$r << N^{1/9}$.

The M(atrix)-model calculation is a perturbative loop expansion
and we must investigate whether its range of validity overlaps with
supergravity. The M(atrix)-loop expansion is an expansion in $N/r^3$ and
we therefore
need $N/r^3 << 1$.
Hence we find
\be
N^{1/3} << r << N^{1/9}
\ee

It follows that
there is no overlap between the M(atrix) regime and
the supergravity regime. There is therefore no reason to expect the
perturbative M(atrix) calculations to agree with supergravity
and we can  expect agreement only when there is a nonrenormalization
theorem to rely on. 

The model example where there is nonrenormalization
is the scattering of two gravitons. At one loop
it can be argued from supersymmetry that the $v^2$ term
vanishes and that the $v^4$ term is not renormalized and agrees
directly with supergravity \cite{dkps}. From the D-particle point of
view this
follows from the force being of the same form for short and long
distances.
The successful two loop calculation in \cite{beckers,bbpt} suggests that
there is
nonrenormalization of terms at least up to order  $v^6$. The discussion
of \cite{dineir} suggests, however, possible problems at higher orders.
Related are the successful one loop
calculations for extended objects \cite{matrixcalculations}
represented by particular background matrices.
An alternative representation for
5-branes is described in \cite{db} through explicit addition of 
hypermultiplets to the M(atrix) theory.
At one loop the supergravity result is again successfully reproduced. 
This is a
parallel to the case of two D-particles, but now it is the $v^2$ which
is 
protected with the force being of the same form for short and long
distances. 

Other cases where the action is modified are orientifolds.
Various such examples were discussed in \cite{ggr}. Three cases
are successful at the one loop level. These are $R/Z_2$
(see \cite{hetliv}), $R^5/Z_2$ and $R^9/Z_2$,
e.g. those cases where you can find protected terms in the expansion
according to the above prescription. 
For the other orientifolds, the situation is not as fortunate. 
In the case of $R^8/Z_2$, 
it is noted in \cite{ggr} that
a suitable term can be found at two loops but with the wrong
N-dependence.
In our view there is no reason to expect agreement. No
nonrenormalisation 
theorem that might protect from higher loop corrections is expected.

In this paper we will investigate graviton scattering in the $R^5/Z^2$ case
\cite{dm96,witten,fs}. 
At one loop, \cite{ggr}, the effective potential contains a piece involving 
an interaction between the graviton and its mirror image of order 
$N^2 v^4/r^7$ and an interaction directly with the five brane charge of 
the orientifold of order $N v^2/r^3$. We will
verify that there is no contribution to the $v^2$ at two loops but  
we do find a term of order $N^2 v^4/r^{10}$. The result is
furthermore shown to be in exact agreement with supergravity, where the
contribution can be seen to be a three graviton effect involving the 
orientifold,
the graviton and the mirror image.

Recently several papers have appeared discussing
three graviton scattering, \cite{ffi,tvr,OY,eg,msw}. 
An explanation of the discrepancy discussed in \cite{arvind} has been
suggested in \cite{tvr}, and the exact agreement between the approaches 
verified in \cite{OY}.
The present work is a further successful test of the M(atrix) model along 
these lines.

The outline of the paper is as follows. In section~\ref{sec:2} we perform
the supergravity calculation. In section~\ref{sec:3} we perform the two loop
M(atrix) model calculation with some details presented in appendices.
Section~\ref{sec:con} contains our conclusions.

\section{The supergravity calculation}
\label{sec:2}

 From the supergravity point of view we need to calculate the scattering
of two gravitons (a source and a probe) in the presence of an object 
with five brane charge. 
This is due to the fact
that the $Z_2$ quotient produces an image for each particle, but also 
the 5-dimensional fixed plane becomes a source for the four form and the 
graviton, \cite{witten}.

The metric and four form corresponding to a five dimensional object charged 
under 
$F_{[4]}$ are 
\beqs
ds^2 &=& H^{-1/3}(dx^{+} dx^- + dx_i dx^i)+ H^{2/3} dy_I dy^I \\
F_{I_1I_2I_3I_4} &=& 3 k \epsilon_{I_1I_2I_3I_4I_5} 
\frac{y^{I_5}}{\rho _\perp ^{5}} ,
\label{eq:bkg}
\eeqs
where $H=1+\frac{k}{\rho _\perp ^3}$, with $k$ a parameter given by the five 
brane charge.
$i= 1...4$, $I=5...9$ and we have defined 
$\rho _\perp ^2=y_Iy^I$ and $\rho _{||} ^2=x_ix^i$. We 
have introduced light-cone coordinates
$x^{\pm}=x_{11}
\pm t$ where $x^+$ plays the role of time $\tau=\frac{1}{2} x^+$. To the
order that we will be working the field strength $F_{[4]}$ will not be important.

The energy momentum tensor of the image graviton can be computed from 
the action for a massless particle
\be
S = \frac{1}{2} \int d\tau \ e^{-1} g_{\mu\nu} \dot{x}^{\mu} \dot{x}^{\nu} ,  
\ee
where $e$ is an arbitrary function of the parameter $\tau$ which we choose to be 
constant.
The energy momentum tensor is then:
\beqs
T_{\mu\nu}(x) &=& \frac{1}{\sqrt{-g}} \frac{\delta S}{\delta g^{\mu\nu}(x)} \\
           &=&  \frac{1}{2\sqrt{-g}}\int d\tau\ e^{-1} \delta(x-x(\tau)) 
               g_{\mu\alpha} g_{\nu\beta} \dot{x}^{\mu} \dot{x}^{\nu} .
\eeqs
Using
\be
p_{-} = e^{-1} g_{-+} \dot{x}^{+} = \frac{N}{R}
\ee
where $R$ is the radius of the compact dimension $x^-$,
one finds
\be
T^{\mu\nu} = p_{-} \delta ^{(9)} (x^i ,y^I ) u^\mu u^\nu ,\label{eq:tmu}
\ee
where $u^\mu$ is the velocity of the image graviton 
(i.e. the source graviton) and is given by
\be
u^+ = 2,\ u^-=-\frac{1}{8} v^2,\ \vec{u} =-\frac{v}{2} \breve{y}^2 ,
\ee
where $\breve{y}^2$ is a unit vector pointing in the $y^2$ direction, i.e. 
$u^\mu$ corresponds to a D-particle moving along $y^2$. 
The impact parameter will be given
by $y^1 =-r$.
The probe will be moving in the opposite direction with
\be
s^+ = 2,\ s^-=-\frac{1}{8} v^2,\ \vec{s} =\frac{v}{2} \breve{y}^2 ,
\ee
and impact parameter $y^1 =r$.
The relative velocity of the probe and source gravitons is $v$.
This means that the image particle produces the Aichelburg-Sexl metric 
\be
g^{AS}_{\mu\nu} = \eta_{\mu\nu} + \zeta ^h_{\mu\nu},
\ee
where
\be
\zeta ^h_{\mu \nu} = \frac{15 N}{2R^2M^9(\rho _\perp ^2 +\rho _{||} 
^2)^{\frac{7}{2}}} 
u_\mu u_\nu \equiv h_0 u_\mu u_\nu ,
\label{eq:ASm}
\ee
after averaging over the $x^-$ direction. Here $M$ is the 11-dimensional
Planck mass and we used the convention $\kappa_{11}^2=16\pi^5/M^9$.

\begin{figure}  
\begin{center}                                                         
\epsfig{file=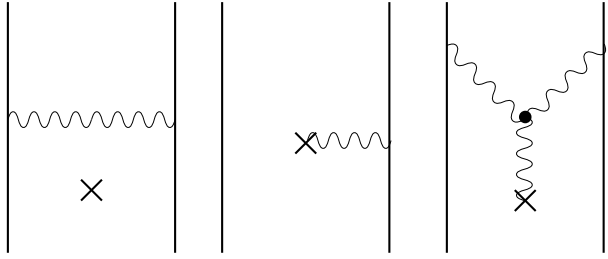,width=8cm}
\caption{Interactions between the graviton and its mirror image, directly with
the orientifold and an interaction involving the graviton, its image and
the orientifold.}
\label{fig:ftdiags}
\end{center}
\end{figure}

The fact that makes the calculation interesting is that the resulting metric
is not a linear superposition of (\ref{eq:bkg}) and (\ref{eq:ASm}). Since
the equation of motion are non-linear there will be corrections that
include a three graviton vertex as depicted in fig.\ref{fig:ftdiags}.
We will  now proceed to evaluate these corrections.

We expand the metric around a Minkowski background according to
\be
g_{\mu \nu} = \eta _{\mu \nu} + h_{\mu \nu} ,
\ee
where the perturbation is given by 
\be
h_{\mu \nu} = \zeta ^h _{\mu \nu} +\zeta ^H _{\mu \nu} + \chi _{\mu \nu} .
\ee
The first two terms are the super imposed unperturbed backgrounds where
the D-particle part is given by
equation (\ref{eq:ASm}) above
while the orientifold part is given by
\be
\zeta ^H _{-+} =-\frac{k}{6\rho_\perp^3} \;\;\; \zeta ^H _{\parallel \parallel} 
=-\frac{k}{3\rho_\perp^3} 
\;\;\; \zeta ^H _{\perp \perp} =\frac{2k}{3\rho_\perp ^3} ,
\ee
and $\chi _{\mu \nu}$ is the genuine three body contribution
that will be obtained below.

The effective action consists of two parts. The D-particle Lagrangian
contributes with
\be
L_D = \frac{N}{R} \left[ \frac{v^2}{8} +\frac{1}{2} h_{\mu \nu} s^\mu s^\nu (1-h_{- \lambda
}s^{\lambda}) \right].
\ee
With our ansatz for the metric and using
e.g. $s^\mu u_\mu = -v^2/2$
we find
\be
L_D = \frac{N}{R} \left[ \frac{v^2}{8} (1+\frac{k}{r^3}) + \frac{1}{8}h_0 v^4
(1+\frac{5k}{6r^3}) 
+ \frac{1}{2} \chi _{\mu \nu} s^\mu s^\nu \right].
\ee
Adding the gravitational Lagrangian given by~\cite{OY}
\be
-\frac{N}{R}\left[ \frac{1}{4} \zeta _{\mu \nu} s^\mu s^\nu 
                  +\frac{1}{3} \chi _{\mu \nu}  s^\mu s^\nu\right] ,
\ee
we obtain
\be
L = \frac{N}{R} \left[ \frac{v^2}{8} (1+\frac{k}{2r^3}) + \frac{1}{16}h_0 v^4
(1+\frac{5k}{3r^3}) 
+ \frac{1}{6} \chi _{\mu \nu} s^\mu s^\nu \right] .
\ee
We must now determine $\chi _{\mu \nu} s^\mu s^\nu$ and to this end we need
to consider the Einstein tensor.
We find a first order piece given by
\be
G^{(1)}_{\mu \nu} = \frac{1}{2} (-\partial ^2 h_{\mu \nu} 
-\partial _\mu \partial _\nu h^\alpha _\alpha +
\partial _\mu \partial _\alpha h^\alpha _\nu +
\partial _\nu \partial _\alpha h^\alpha _\mu)  ,
\ee
and a second order piece given by
\beqs
G^{(2)}_{\mu \nu} = -\frac{1}{2} h_{\alpha \beta}( \partial ^\alpha \partial 
_\nu h^\beta _\mu
+\partial ^\alpha \partial _\mu h^\beta _\nu 
-\partial ^\alpha \partial ^\beta h_{\mu \nu}  
-\partial _\mu \partial _\nu h^{\alpha \beta} )
\nonumber  \\
+\frac{1}{2} \partial _\alpha h_{\beta \nu}\partial ^\alpha h_\mu ^\beta
-\frac{1}{2} \partial _\alpha h_{\beta \nu}\partial ^\beta h_\mu ^\alpha
+\frac{1}{4} \partial _\nu h_{\alpha \beta}\partial _\mu h^{\alpha \beta} 
\nonumber \\
-\frac{1}{2} (\partial _\mu h^\alpha _\nu + \partial _\nu h^\alpha _\mu
-\partial ^\alpha  h_{\mu \nu}) 
(\partial ^\beta  h_{\alpha \beta}-\frac{1}{2} \partial _\nu h^\alpha _\alpha ) 
\nonumber \\
-\frac{1}{2} h_{\mu \nu}(-\partial ^2 h^\alpha _\alpha +\partial _\alpha 
\partial _\beta h^{\alpha \beta} ) 
-\frac{1}{2} \eta _{\mu \nu} R .
\eeqs
We impose the gauge choice
\be
\partial ^\mu (\chi _{\mu \nu}  -\frac{1}{2} \eta _{\mu \nu} \chi
^\lambda _\lambda ) =0 ,
\ee
and obtain  $-\frac{1}{2} \partial ^2 \chi _{\mu \nu}$ from the
first order term.
The same gauge condition is also obeyed by the 
unperturbed background $\zeta _{\mu \nu}$.
However, as opposed
to \cite{OY} the background is not traceless. $\zeta ^H_{\mu \nu}$  has a 
trace given by 
$\zeta^{H\lambda} _\lambda  =\frac{4k}{3r^3}$. 
The relevant source terms obtained
from the second order piece contain one $\zeta ^h$ and one $\zeta ^H$, note that
the third line vanishes due to the gauge condition.
Furthermore, the energy momentum tensor (\ref{eq:tmu}) provides 
a source term for $\chi _{\mu \nu}$ when the
indices are lowered by the orientifold metric. Following \cite{OY} we find
the relevant contribution to be
\be
\kappa _{11} ^2 (T_\nu ^\alpha \zeta ^H _{\alpha \mu} 
+T_\mu ^\alpha \zeta ^H _{\alpha \nu}) .
\ee
This simple energy momentum tensor where all the non-linearities are coming from
the lowering of the indices is crucial for obtaining the M(atrix) 
model result both for
\cite{OY} and in our case. 
As explained in \cite{tvr} further non-linearities are absorbed into
the quantized lightlike momentum.
We put 
\be
\chi = \chi _{\mu \nu} s^\mu s^\nu ,
\ee
and for convenience we redefine 
\be
h _{\mu \nu} = H^{2/3} \tilde{h} _{\mu \nu}  .
\ee
The resulting equation for $\tilde{\chi}$ is
\beqs
\Delta \tilde{\chi} +
\frac{v^4}{4} \bigg( \frac{k}{\rho ^3} \Delta _{||} h_0 +\nabla h_0 \cdot  
\nabla \frac{k}{\rho ^3}
+\frac{2\kappa _{11}^2k}{3\rho ^3} p_- \delta ^{(9)} (x^i,y^I) \bigg) \nonumber 
\\
+
\frac{v^4}{4} \bigg( \frac{k}{\rho ^3} \partial _y^2 h_0 -\frac{1}{6} \partial 
_y h_0  
\partial _y \frac{k}{\rho ^3}
-\frac{1}{3} h_0 \partial _y^2 \frac{k}{\rho ^3} \label{eq:chi}
 \bigg) =0
\eeqs
where $y=y^2$ is the direction along the motion of the graviton.
We need not solve the equation in general since we are interested only in 
the value of $\chi$ at the position of the particle. It is in fact
enough to consider the case
$(y^1=r,y^2=\cdots y^4 = x^i =0)$. This
can be obtained using the 9-dimensional Green function. 
For convenience we define $\rho _\perp ^2 = \rho ^2 + z^2$ where $z=y^1$.
In particular we find that the first of the source terms 
in equation (\ref{eq:chi}) leads to the integral
$$
\frac{225}{16} \int
_{-\infty}^{\infty} dz \int _{0}^{\infty} 
d\rho_\perp d\rho_\parallel \rho_\perp^3 \rho _\parallel^3 
\frac{1}{ ((z-r)^2+\rho _\perp^2 
+\rho _\parallel^2 )^{7/2}} 
$$
\be
\times \frac{1}{(z^2+\rho _\perp^2 )^{3/2}}  \Delta _{\parallel}
\frac{1}{ ((z+r)^2+\rho _\perp^2  +\rho _\parallel ^2 )^{7/2}} =
-\frac{105}{256r ^{10}} = -\frac{7k}{r ^3} h_0 .
\ee
The integral has been
evaluated using a change of variables according to 
$r_1^2 =(z-r)^2+\rho _\perp^2 +\rho _\parallel^2 $, 
$r_2^2 =(z+r)^2+\rho _\perp ^2+\rho _\parallel^2 $, and $r_3^2 =z^2+\rho 
_\perp^2$.
A similar integral gives $\frac{8k}{\rho ^3} h_0$ for the second term
of the first line while
the last term of the first line immediately
gives $\frac{k}{3\rho^3} h_0$. It can be verified that the 
rest of the terms
do not contribute to the particular
scattering amplitude that we are interested in. We can then conclude
that
\be
\tilde{\chi} = \frac{v^4}{3}h_0 \frac{k}{r^3}
\ee
and therefore
\be
\chi = \frac{v^4}{2}h_0 \frac{k}{r^3}   .
\ee
This allows us to obtain
\beqs
L &=& \frac{N}{R} \left[ \frac{v^2}{8} (1+\frac{k}{2r^3}) +\frac{1}{16}h_0 v^4
(1+\frac{3k}{r^3})\right] \nonumber \\
  &=& \frac{N}{R} \frac{v^2}{8} (1+\frac{k}{2r^3}) + 
\frac{N^2}{R^3M^9} \frac{15}{32}\frac{v^4}{(2r)^7} (1+\frac{3k}{r^3}) 
\label{eq:supres0} .
\eeqs
This is however not the whole story. We must also add the identical
contribution from the image graviton action and the action of the
five brane charged object (including gravitational actions). 
The latter is easily deduced without
any further calculation. The integrated force on the orientifold must be
equal and opposite the one from the two gravitons, it will therefore
be precisely equal to the sum of the contributions proportional to $k$
in the action of the two gravitons. This effectively doubles the value
of $k$ and we obtain
\beqs
L _{tot}
&=& 2\bigg(\frac{N}{R} \frac{v^2}{8} (1+\frac{k}{r^3}) + 
\frac{15}{32}\frac{N^2}{R^3M^9} \frac{v^4}{(2r)^7} 
(1+\frac{6k}{r^3}) \bigg) \nonumber\\
&=& 2\bigg( \frac{N}{R} \frac{v^2}{8} (1-\frac{1}{2(Mr)^3}) + 
\frac{15}{32}\frac{N^2}{R^3M^9} \frac{v^4}{(2r)^7} (1-\frac{3}{(Mr)^3}) \bigg) 
\label{eq:supres} ,
\eeqs
where we have put $k=-1/(2M^3)$. Note that the value for a five brane is
$k=1/(2M^3)$ as follows from e.g. \cite{db} with the convention that
$\kappa_{11}^2=16\pi^5/M^9$. As explained in
\cite{ggr} the value for the orientifold comes from working on the
covering space. At order $v^2/r^3$ the effect of
the gravitational action is to cancel the five brane contribution
explaining in the spirit of \cite{OY} why previous calculations,
e.g. \cite{dkps},
where the gravitational action (and five brane action) have been neglected,
nevertheless give the right answer.

\section{The M(atrix) model calculation}
\label{sec:3}
 
We now turn to the M(atrix) theory description of the $R^5/Z_2$ orbifold with
the aim of reproducing the supergravity result (\ref{eq:supres}). 
To derive the action we start with a $U(2N)$ 0+1 dimensional 
supersymmetric Yang-Mills theory on the covering space with fields
$X^i, \:i=1,\ldots,9$ in the adjoint representation, which describe
$N$ D0-branes along with their mirror images.  The fields are split into 
$X_{\perp}$
and $X_\parallel$ corresponding to directions parallel and transverse to the 
orientifold plane. Keeping only states that are invariant
under the combined action of orientation reversal and
space-time reflection gives rise to the projection condition \cite{hetliv}
\bea
X_\parallel  &=& M X^{\rm T}_\parallel  M^{-1}, \nonumber \\
X_\perp  &=& -  M X^{\rm T}_\perp M^{-1}, \nonumber \\
\Theta &=& \Gamma_\perp \, M  \Theta^{\rm T} M^{-1},
\hskip1.5cm \Gamma_\perp = \Gamma_5 \cdots \Gamma_9.
\label{eq:projcond}
\eea
These equations restricts the gauge group for the remaining states
to be either $SO(2N)$ or $USp(2N)$, but 
consistency conditions enforce the choice $USp(2N)$ \cite{fs,kr2}. The action 
is
given by \cite{kr2} 
\beqs
        S&=&\frac{1}{2}\int dt\;\mathrm{Tr}\Bigg\{\frac{1}{2R}
        (D_t X_i)^2 +  \frac{1}{2R} (D_t Y_I)^2  + {R \over 4} [X_i,X_j] ^2
        + {R \over 4} [Y_I,Y_J] ^2 \nonumber \\
        &&+ {R \over 2} [Y_I,X_j] ^2 
        +  S_a D_t S_a + S_\da D_t S_\da  
        +2 i R X_i \sigma^i_{a\da} \{S_a, S_\da\} \nonumber \\
        &&-S_a \gamma^I [Y_I, S_a] 
        +S_\da \gamma^I [Y_I, S_\da] 
        \Bigg\},
         \label{action1}
\eeqs
where $R$ is the eleven dimensional radius and we have put the eleven 
dimensional 
Planck length $l_p=1$. The indices take values $i=1,\ldots,4$ and $I=5,\ldots,9$
and $D_t = \partial_t -i [A_0, \cdot]$. An extra factor of $1/2$ has been 
included
in front of the action in account for the fact that our system does not include 
the action of the image. This is clearly seen from the kinetic term which 
must be ${N \over 2R} ({v \over 2})^2$. This is equivalent to take the coupling constant 
$g=4R$ instead of $g=2R$ in the notation of \cite{bbpt}.
 
The four $X_i$ are in the antisymmetric representation of $Sp(N)$ and
corresponds to directions parallel to the orientifold, the fifth one is
the longitudinal direction of the M(atrix)-model. The $Y_I$ with $I=5,...,9$
give the transversal coordinates. The fermionic fields are $S_a$ in the adjoint
representation and $S_{\dot a}$ in the antisymmetric representation.
We will perform a loop calculation using the background field method,
As we will consider scattering in a direction transverse to the fixed 
plane,
the background field $B_I$ only has non-zero components $B_I$, $I=5,\ldots,9$.
To the Lagrangian we add ghosts and a gauge fixing term
\be
L_{gf}=-{1 \over 2} \left( \partial_t A_0 - i[B^I ,Y_I] \right)^2  .
\ee
Expanding around the background 
\be
Y_I=B_I+A_I,
\ee
where $B_I$ is chosen to satisfy the
equations of motion we obtain
\beqs
L&=&{1 \over 2} \mathrm{Tr}   \Bigg\{ \dot A_I ^2 +  \dot X_i ^2 -  \dot A_0^2
- 4 i\dot B_I [A_0,B^I] - [A_0,B_I]^2 + [B_I,A_J]^2 \nonumber \\
&&+[B_I,X_j]^2 +[B^I,A_I]^2 + [B_I,A_J][A^I,B^J]+ [B_I,B_J][A^I,A^J]  \nonumber 
\\
&&- 2 i\dot A_I [A_0,A^I] - 2 i \dot X_i [A_0,X^i] - 2 [A_0,B_I][A_0,A^I] 
 +2 [B_I,A^J][A^I,A^J]  \nonumber \\
&& +2 [B_I,X_i][A^I,X^i]  -[A_0,X_i]^2  -[A_0,A_I]^2  + {1 \over 2} [A_I,A_J]^2 
\nonumber \\
&&+ {1 \over 2} [X_i,X_j]^2
+  [A_I,X_j]^2 \nonumber \\
&&+\dot C^* C + [C^*,B^I][B_I,C] + [C^*,B^I][A_I,C]+ i \dot C^*[C,A_0] \nonumber 
\\
&&+ S_a \dot S_a + S_\da \dot S_\da    - i
S_a[A_0, S_a] - i S_\da[A_0, S_\da] - S_a\gamma _I [A_I, S_a] \nonumber \\
&&-  S_a\gamma _I [B_I, S_a] + S_\da \gamma^I [A_I, S_\da] + S_\da \gamma^I 
[B_I, S_\da]
++ 2X_i \sigma^i_{a\da} \{S_a, S_\da\}  \Bigg\},
\label{action2}
\eeqs
where we have performed
the usual rescaling of fields~\cite{ggr}. The calculation 
will be 
performed in Euclidean time so we introduce $\tau=it$ and $A_\tau=-iA_0$.

%----------------------------------------------------------------------------
%____________________________________________________________________________

The background that we will use is:
\be
B_9=\frac{v t}{2} \sigma_3\otimes 1_{N\times N},\;\;
B_8= b \sigma_3 \otimes 1_{N\times N}  ,
\ee
where $\sigma_3$ is a Pauli matrix and $1_{N\times N}$ is the identity matrix of 
$N\times N$.
This corresponds to a particle composed of $N$ D0-branes moving in a direction 
transverse
to the orientifold with an impact parameter $b$ and velocity $v/2$ relative to
the fixed point. (The distance between the particle and its mirror image is thus
$\sqrt{(2b)^2+(vt)^2}$ and the relative velocity is $v$).

Expanding around the background we obtain terms quadratic in the fluctuations 
which 
determine the propagators and cubic and quartic terms which give vertices with 
three and four 
legs respectively.  
The propagators are given in Appendix B and correspond to massless 
and massive fluctuations. 
Massless ones associated with strings stretching between D0-branes in the 
particle 
and massive
to strings stretching between a D0-brane in the particle and another in the 
image.
In terms of matrices, these are described by block diagonal and off-diagonal 
matrices
respectively.
The matrices $X_{\mu=5\ldots 9}$ are in the adjoint representation which is 
$N(2N+1)$-dimensional. 
Of them, $N^2+N$ become
massive. On the other hand, matrices $X_{\mu=1\ldots 4}$ are in the two index 
antisymmetric representation 
having $2N^2-N-1$ fields out of which $N^2-N$ aquire a mass. In total we get the 
following 
massive fields:

\begin{itemize}
\item $8 N^2$ bosons of mass $m^2=(2r)^2$.
\item $N^2 + N$ bosons of mass $m^2=(2r)^2+2v$.
\item $N^2 + N$ bosons of mass $m^2=(2r)^2-2v$.
\item $N^2 + N$ complex ghosts of mass $m^2=(2r)^2$.
\item $4 N^2$ fermions of mass $m^2=(2r)^2+v$.
\item $4 N^2$ fermions of mass $m^2=(2r)^2-v$.
\end{itemize}
where $r = \sqrt{b^2+(vt/2)^2}$ is the distance between the particle and the 
orientifold.
Integrating out massive fields at one loop we obtain the potential
\be
V = N \frac{v^2}{2 (2r)^3} - \frac{v^4}{(2r)^7} \left( \frac{15}{32} N^2 + \frac 
{5}{16} 
N\right) . 
\ee
We note that the $N^2$ contribution is one half of the one-loop potential
between two D0-branes in uncompactified space \cite{dkps}, while for $N=1$ 
we obtain the result of \cite{kr2}.

\begin{figure}  
\begin{center} 
\epsfig{file=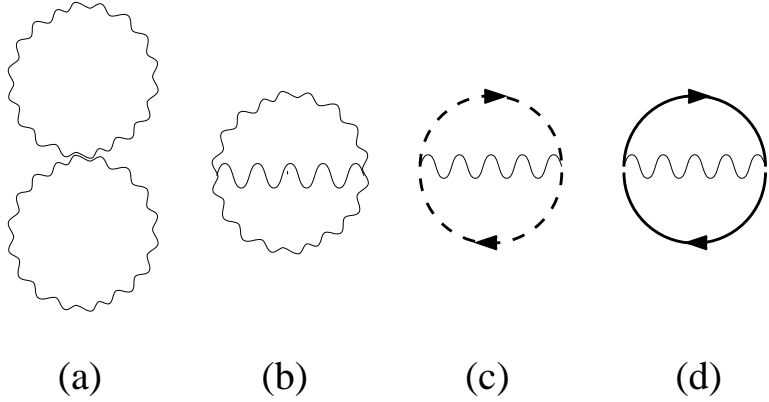}
\caption{Two loop Feynman diagrams.}
\label{feynman}
\end{center}
\end{figure}

To obtain the two-loop correction we must evaluate vacuum diagrams (see Fig.
\ref{feynman}) 
using
the cubic and quadratic vertices. The resulting expression for the phase shift 
is 
expanded in powers of the velocity . Finally we calculate a corresponding 
potential.

In the following we give the final result for each diagram whereas intermediate 
expressions
can be found in Appendix B. Some formulas used in the computation of
group theory factors are given in Appendix A.

\begin{description}
\item{a) Diagram with a quartic vertex}

\beqsn
(a) &=& \left( \frac{45}{16} \frac{1}{(2r)^2} + \frac{135}{128} 
\frac{v^2}{(2r)^6} + 
\frac{7551}{2048} \frac{v^4}{(2r)^{10}} + \frac{177405}{114688} 
\frac{v^6}{(2r)^{14}} + \ldots 
\right) N^3 \\
    &+& \left( \frac{9}{8} \frac{1}{(2r)^2} + \frac{123}{64} \frac{v^2}{(2r)^6} 
+ 
\frac{20799}{5120} \frac{v^4}{(2r)^{10}} + \frac{287481}{57344} 
\frac{v^6}{(2r)^{14}} + \ldots 
\right) N^2 \\
    &+& \left( -\frac{3}{16} \frac{1}{(2r)^2} + \frac{55}{128} 
\frac{v^2}{(2r)^6} + 
\frac{9387}{10240} \frac{v^4}{(2r)^{10}} + \frac{220685}{114688} 
\frac{v^6}{(2r)^{14}} + \ldots 
\right) N .
\eeqsn

\item{ b0) Two bosonic cubic vertices without time derivatives}
 
 \beqsn
(b0) &=& \left( \frac{27}{64} \frac{1}{(2r)^2} + \frac{1}{4} \frac{v^2}{(2r)^6} 
+ 
\frac{132417}{40960} \frac{v^4}{(2r)^{10}} - \frac{3261963}{917504} 
\frac{v^6}{(2r)^{14}} + 
\ldots \right) N^3 \\
    &+& \left( \frac{7}{32} \frac{1}{(2r)^2} + \frac{251}{192} 
\frac{v^2}{(2r)^6} + 
\frac{85077}{20480} \frac{v^4}{(2r)^{10}} + \frac{920753}{196608} 
\frac{v^6}{(2r)^{14}} + \ldots 
\right) N^2 \\
    &+& \left( \frac{19}{64} \frac{1}{(2r)^2} + \frac{67}{192} 
\frac{v^2}{(2r)^6} + 
\frac{120009}{40960} \frac{v^4}{(2r)^{10}} - \frac{5875921}{2752512} 
\frac{v^6}{(2r)^{14}} + 
\ldots \right) N .
\eeqsn
 
 \item{ b1) Two bosonic cubic vertices with one time derivative}
 
 \beqsn
(b1) &=& \left( - \frac{27}{64} \frac{v^2}{(2r)^6} - \frac{9}{32} 
\frac{v^4}{(2r)^{10}} + 
\frac{13941}{16384} \frac{v^6}{(2r)^{14}} + \ldots \right) N^3 \\
    &+& \left( - \frac{3}{8} \frac{v^2}{(2r)^6} + \frac{27}{128} 
\frac{v^4}{(2r)^{10}} + 
\frac{6813}{4096} \frac{v^6}{(2r)^{14}} + \ldots \right) N^2 \\
    &+& \left( \frac{3}{64} \frac{v^2}{(2r)^6} + \frac{63}{128} 
\frac{v^4}{(2r)^{10}} + 
\frac{13311}{16384} \frac{v^6}{(2r)^{14}} + \ldots \right) N  .
\eeqsn

 \item{ b2) Two bosonic cubic vertices with two time derivatives}
 
 \beqsn
(b2) &=& \left( -\frac{81}{64} \frac{1}{(2r)^2} - \frac{1379}{512} 
\frac{v^2}{(2r)^6} - 
\frac{188631}{40960} \frac{v^4}{(2r)^{10}} - \frac{2366913}{458752} 
\frac{v^6}{(2r)^{14}} + 
\ldots \right) N^3 \\
    &+& \left( -\frac{45}{32} \frac{1}{(2r)^2} - \frac{831}{256} 
\frac{v^2}{(2r)^6} - 
\frac{24351}{4096} \frac{v^4}{(2r)^{10}} - \frac{1677897}{229376} 
\frac{v^6}{(2r)^{14}} + \ldots 
\right) N^2 \\
    &+& \left( -\frac{9}{64} \frac{1}{(2r)^2} - \frac{283}{512} 
\frac{v^2}{(2r)^6} - 
\frac{54879}{40960} \frac{v^4}{(2r)^{10}} - \frac{988881}{458752} 
\frac{v^6}{(2r)^{14}} + \ldots 
\right) N .
\eeqsn

 \item{ c0) Two ghost vertices without time derivatives}
 
 \beqsn
(c0) &=& \left( -\frac{1}{64} \frac{1}{(2r)^2} + \frac{17}{768} 
\frac{v^2}{(2r)^6} - 
\frac{2571}{40960} \frac{v^4}{(2r)^{10}} - \frac{892261}{2752512} 
\frac{v^6}{(2r)^{14}} + \ldots 
\right) N^3 \\
    &+& \left( -\frac{1}{32} \frac{1}{(2r)^2} + \frac{17}{384} 
\frac{v^2}{(2r)^6} - 
\frac{2571}{20480} \frac{v^4}{(2r)^{10}} + \frac{892261}{1376256} 
\frac{v^6}{(2r)^{14}} + \ldots 
\right) N^2 \\
    &+& \left( -\frac{1}{64} \frac{1}{(2r)^2} + \frac{17}{168} 
\frac{v^2}{(2r)^6} - 
\frac{2571}{40960} \frac{v^4}{(2r)^{10}} + \frac{892261}{2752512} 
\frac{v^6}{(2r)^{14}} + \ldots 
\right) N .
\eeqsn
 
 \item{ c1) Two ghost vertices with one time derivative}
 
 \beqsn
(c1) &=& \left(  \frac{1}{32} \frac{v^2}{(2r)^6} + \frac{9}{256} 
\frac{v^4}{(2r)^{10}} + 
\frac{231}{4096} \frac{v^6}{(2r)^{14}} + \ldots \right) N^3 \\
    &+& \left(   \frac{1}{16} \frac{v^2}{(2r)^6} + \frac{9}{128} 
\frac{v^4}{(2r)^{10}} + 
\frac{231}{2048} \frac{v^6}{(2r)^{14}} + \ldots \right) N^2 \\
    &+& \left( \frac{1}{32} \frac{v^2}{(2r)^6} + \frac{9}{256} 
\frac{v^4}{(2r)^{10}} + 
\frac{231}{4096} \frac{v^6}{(2r)^{14}} + \ldots \right) N .
\eeqsn

 \item{ c2) Two ghost vertices with two time derivatives}
 
 \beqsn
(c2) &=& \left( \frac{3}{64} \frac{1}{(2r)^2} + \frac{73}{512} 
\frac{v^2}{(2r)^6} + 
\frac{7893}{40960} \frac{v^4}{(2r)^{10}} + \frac{107251}{458752} 
\frac{v^6}{(2r)^{14}} + \ldots 
\right) N^3 \\
    &+& \left( \frac{3}{32} \frac{1}{(2r)^2} + \frac{73}{256} \frac{v^2}{(2r)^6} 
+ 
\frac{7893}{20480} \frac{v^4}{(2r)^{10}} + \frac{107251}{229376} 
\frac{v^6}{(2r)^{14}} + \ldots 
\right) N^2 \\
    &+& \left( \frac{3}{64} \frac{1}{(2r)^2} + \frac{73}{512} \frac{v^2}{(2r)^6} 
+ 
\frac{7893}{40960} \frac{v^4}{(2r)^{10}} + \frac{107251}{458752} 
\frac{v^6}{(2r)^{14}} + \ldots 
\right) N .
\eeqsn

 \item{ d) Two fermionic vertices }
 
 \beqsn
(d) &=& \left( -2 \frac{1}{(2r)^2} + \frac{155}{96} \frac{v^2}{(2r)^6} - 
\frac{1407}{640} 
\frac{v^4}{(2r)^{10}} + \frac{1132235}{172032} \frac{v^6}{(2r)^{14}} + \ldots 
\right) N^3 \\
    &+& \left( -\frac{15}{32} \frac{v^2}{(2r)^6} + \frac{135}{128} 
\frac{v^4}{(2r)^{10}} + 
\frac{42705}{81922} \frac{v^6}{(2r)^{14}} + \ldots \right) N ,
\eeqsn
where we can see that as in the one loop case, the fermions do not contribute to 
the $N^2$ term.
\end{description}
Summing up all the bosonic plus ghost diagrams we obtain

\beqsn
(\mbox{bos. + gh.}) &=& \left(  2 \frac{1}{(2r)^2} - \frac{155}{96} 
\frac{v^2}{(2r)^6} + 
\frac{1407}{640} \frac{v^4}{(2r)^{10}} + \frac{981035}{172032} 
\frac{v^6}{(2r)^{14}} + \ldots 
\right) N^3 \\
    &+& \left(   \frac{45}{16} \frac{v^4}{(2r)^{10}} + \frac{675}{128} 
\frac{v^6}{(2r)^{14}} + 
\ldots \right) N^2 \\
    &+& \left( \frac{15}{32} \frac{v^2}{(2r)^6} + \frac{405}{128} 
\frac{v^4}{(2r)^{10}} + 
\frac{7695}{8192} \frac{v^6}{(2r)^{14}} + \ldots \right) N .
\eeqsn
Adding the fermionic contribution the final result is obtained:

\beqs
(\mbox{bos. + gh. + ferm.}) &=& \left(  \frac{225}{256} \frac{v^6}{(2r)^{14}} + 
\ldots \right) 
N^3 \nonumber\\
    &+& \left(   \frac{45}{16} \frac{v^4}{(2r)^{10}} + \frac{675}{128} 
\frac{v^6}{(2r)^{14}} + 
\ldots \right) N^2 \\
    &+& \left( \frac{135}{32} \frac{v^4}{(2r)^{10}} - \frac{1575}{256} 
\frac{v^6}{(2r)^{14}} + 
\ldots \right) N .\nonumber
\label{eq:diagrams}
\eeqs

The first thing to note is that there is no $v^2$ term at any order of $N$ 
showing that the
metric is not changed at two-loops. This is in agreement with the 
non-renormalization theorem proved in \cite{de}, see also \cite{pss1,pss2}.
To compare the non-vanishing terms with the 
supergravity 
result it is necessary to include the coupling constant $g = 4 R$ as discussed 
previously.
We must also return to Minkowski space by putting $v \rightarrow iv$.
Powers of $RM^3$, where $M$ is the eleven-dimensional Planck mass,
are also restored using dimensional analysis with the result
\beqs
 L &=& \frac{N}{R} \frac{v^2}{8} - \frac{N}{RM^3}\frac{v^2}{2(2r)^3} + 
\frac{15}{32} \frac{N^2}{R^3M^9}\frac{v^4}{(2r)^7} \nonumber\\ 
 && +\frac{225}{64} \frac{N^3}{R^5M^{18}}\frac{v^6}{(2r)^{14}} 
- \frac{45}{4}\frac{N^2}{R^3M^{12}} \frac{v^4}{(2r)^{10}} +\ldots
 \label{eq:potential}
\eeqs
which includes zero, one and two loop contributions and for each power of the 
velocity, only the 
leading order in $N$ is written.

In a supergravity calculation, the background brings in powers of the 
orientifold charge which is 
proportional to $1/M^3$. Hence the lowest order $1/M$ must represent the 
interaction
between the particle and its image. This interaction potential follows from the 
results of \cite{beckers,bbpt} and 
is given by
\be
V = -\frac{15}{16} \frac{N^2}{R^3M^9} \frac{v^4}{(2r)^7} - \frac{225}{32} 
\frac{N^3}{R^5M^{18}} 
\frac{v^6}{(2r)^{14}} + { O}(\frac{v^8}{r^{21}}).
\ee
This is exactly twice what is included in expression (\ref{eq:potential}) 
in correspondence
with the fact that in (\ref{eq:potential}) only the action of the particle is 
considered and not 
that of the image. On the other hand the agreement is not surprising since it 
can be checked 
diagram by diagram because both calculations are similar. 
Of more interest in our case is the subleading order in $1/M$. The $v^2$ terms 
are
\be
\frac{1}{2}\frac{N}{R}\left(\frac{v}{2}\right)^2
\left(1-\frac{1}{2(Mr)^3}\right) .
\ee
This is the one loop result
that was obtained in \cite{ggr,kr2}.

The $v^4$ term is
\be
 V = -\frac{15}{32} \frac{N^2}{R^3M^9}\frac{v^4}{(2r)^7} 
   (1-\frac{3}{(Mr)^3})  ,
\ee
in precise agreement with the supergravity result of section~\ref{sec:2}. 
The overall factor of two is due to the fact that the action 
(\ref{eq:supres}) includes the image as a separate particle.

\section{Conclusions}
\label{sec:con}

In this paper we have found yet another successful application of the
perturbative M(atrix) model where from the supergravity point of view we are
considering a three graviton effect.
We have one graviton each from the D-particle, the mirror image and the
orientifold meeting in a vertex. It is
interesting to note the agreement in view of the latest developments in the 
study
of three graviton scattering. A natural extension of this work is to consider 
graviton scattering in the presence of a five brane. Two gravitons and one five 
brane
allows more general kinematic configurations than the one studied in this paper.

It is tempting to conjecture an agreement also at higher loops for the
$v^4$ term. This would imply a vanishing of all $N^k v^4$ terms for
$k>2$ above one loop. Furthermore, an extension of the calculation in
section~\ref{sec:2} provides the supergravity prediction for the $N^2 v^4$ term
at any loop.

\section*{Acknowledgements}

The work of U.D. was supported by Swedish Natural Science Research Council (NFR)
and that of M.K. by The Swedish Foundation for International Cooperation in 
Research and Higher Education (STINT).

\renewcommand{\theequation}{\Alph{section}.\arabic{equation}}
\section*{Appendix A}
\setcounter{section}{1}
\setcounter{equation}{0}

To compute the two loop effective action it is necessary to evaluate the 
diagrams 
of 
fig. 2 for the $Sp(N)$ gauge theory. This is a straight forward task which 
proceeds along the same lines of that of \cite{beckers,bbpt}.
In this appendix we will only give some formulas to deal with the $Sp(N)$ 
combinatorics as well as the representation of Dirac matrices we have used.
The representations of $Sp(N)$ under which the matrices $X_{\mu}$ transform 
in the matrix model calculation
are the adjoint and the two-index antisymmetric ones depending on the value of 
$\mu$. 
A convenient set of matrices in which $X_{\mu}$ can be expanded is,   
for the adjoint:
\be
\begin{array}{ccccccc}
\frac{1}{2} \sigma_3 \otimes \lambda_j      &,& 
\frac{1}{2\sqrt{2}} \sigma_3 \otimes S_{ij} &,& 
\frac{1}{2\sqrt{2}} 1_{2\times 2} \otimes A_{ij}, &&\\ &&&&&& \\
\frac{1}{2} \sigma_1 \otimes \lambda_j      &,& 
\frac{1}{2} \sigma_2 \otimes \lambda_j      &,&
\frac{1}{2\sqrt{2}} \sigma_1 \otimes S_{ij} &,&
\frac{1}{2\sqrt{2}} \sigma_2 \otimes S_{ij} ,
\end{array}
\ee
and for two-index antisymmetric representation
\be
\begin{array}{ccccc}
\frac{1}{2} \sigma_3 \otimes \lambda_{j\neq N}      &,& 
\frac{1}{2\sqrt{2}} 1_{2\times 2} \otimes S_{ij} &,& 
\frac{1}{2\sqrt{2}} \sigma_3 \otimes A_{ij}, \\ &&&& \\
\frac{1}{2\sqrt{2}} \sigma_1 \otimes A_{ij} &,&
\frac{1}{2\sqrt{2}} \sigma_2 \otimes A_{ij},
\end{array}
\ee
where $S_{ij}$ and $A_{ij}$ are symmetric and antisymmetric $N\times N$ matrices
given by
\beqs
\left(S_{ij}\right)_{kl} &=& \delta_{ik} \delta_{jl} + \delta_{il} \delta_{jk}, 
\\
\left(A_{ij}\right)_{kl} &=& \delta_{ik} \delta_{jl} - \delta_{il} \delta_{jk}, 
\\
\eeqs
and $\lambda_j$ are diagonal matrices given by 
\beqs
\lambda_{j\neq N} &=& 
\frac{1}{\sqrt{j(j+1)}}\mbox{diag}(\underbrace{1,1,1,\ldots}_{j},-j,0,\ldots,0),
\\
\lambda_N &=&  \frac{1}{\sqrt{N}} 1_{N\times N}.
\eeqs
In both representations the matrices of the first row are block diagonal and 
represent massless strings stretched between the particles whereas the ones in 
the
second row represent strings stretched between a particle and an image.
Since all the above matrices are of the form 
$\frac{1}{\sqrt{2}}\sigma_p \otimes T_a$ with $p=0,1,2,3$, 
($\sigma_0=1_{2\times 2}$) and $T_a$ an $U(N)$ generator normalized to 
$TrT_aT_b=\frac{1}{2}\delta_{ab}$, the expressions for the diagrams can be 
written
in terms of $f_{abc}$ and $d_{abc}$ defined through
\beqs
{[} T_a,T_b ] &=& i f_{abc} T_c ,\\
 \{ T_a,T_b \} &=& d_{abc} T_c .
\eeqs
If one uses this procedure then the following formulas are useful
\beqs
\sum_{a_Ab_Ac_A} \left(f_{abc}\right)^2 &=& 
\frac{1}{4} N^3 - \frac{3}{4} N^2 + \frac{N}{2} ,\\
\sum_{a_Sb_Ac_A} \left(f_{abc}\right)^2 &=& 
\frac{1}{4} N^3 + \frac{1}{4} N^2 - \frac{N}{2} ,\\
\sum_{a_Sb_Sc_S} \left(d_{abc}\right)^2 &=& 
\frac{1}{4} N^3 + \frac{3}{4} N^2 + N ,\\
\sum_{a_Sb_Ac_A} \left(d_{abc}\right)^2 &=& 
\frac{1}{4} N^3 - \frac{1}{4} N^2 ,\\
\eeqs 
where the subscript $S$ or $A$ on an index indicates that it is summed
only over symmetric or antisymmetric generators respectively.

The fermions transform under the same representations and the above 
relations also apply. For the Dirac matrices we have used the following
representation:
\be
\begin{array}{ccccccccccc}
\Gamma_1 &=& \epsilon\otimes\epsilon\otimes\epsilon\otimes\epsilon &\ \ & 
\Gamma_2 &=& \epsilon\otimes\epsilon\otimes 1\otimes\sigma_3 &\ \ &
\Gamma_3 &=& \epsilon\otimes\epsilon\otimes 1\otimes\sigma_1\\ 
\Gamma_4 &=& \sigma_1\otimes\epsilon\otimes\epsilon\otimes 1 &&
\Gamma_5 &=& \sigma_3\otimes\epsilon\otimes\epsilon\otimes 1 &&
\Gamma_6 &=& 1\otimes\epsilon\otimes\sigma_3\otimes\epsilon \\
\Gamma_7 &=& 1\otimes\epsilon\otimes\sigma_1\otimes\epsilon &&
\Gamma_8 &=& 1\otimes\sigma_3\otimes 1\otimes 1               &&
\Gamma_9 &=& 1\otimes\sigma_1\otimes 1\otimes 1 ,
\end{array}
\ee
where $\epsilon = i\sigma_2$ and $\sigma_{1,2,3}$ are the Pauli matrices.
The matrix $\Gamma_{\perp}$ used in (\ref{eq:projcond}) is then:
\be
\Gamma_{\perp} = \Gamma_5\ldots \Gamma_9 = \sigma_3\otimes 1\otimes 1 \otimes 1 ,
\ee
and so the first eight components of $\Theta$ transform in the two-index 
antisymmetric 
and the remaining ones in the adjoint. Using the previous formulas the result 
for
the two loop effective action can be computed resulting in (\ref{eq:potential}). 

\begin{figure}  
\begin{center}                                                         
\epsfig{file=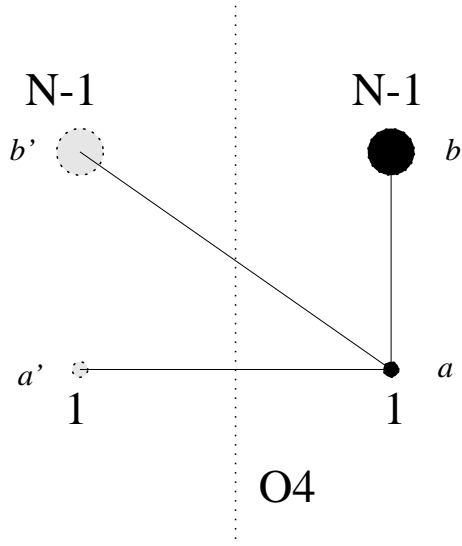,width=6 cm}                                               
     
\caption{String states for the Sp(N) system. We have split the D0-branes
into groups of 1 and $N-1$.}
\label{strings1}
\end{center}
\end{figure}

\begin{figure}  
\begin{center}                                                         
\epsfig{file=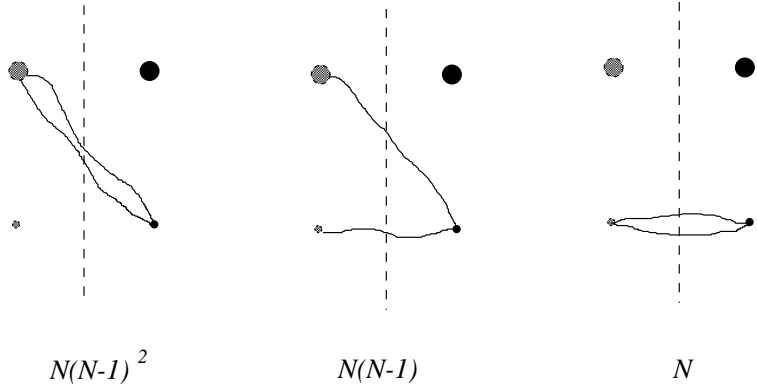,width=10 cm}                                               
     
\caption{The $N$-dependence from different configurations of two massive 
strings.}
\label{strings2}
\end{center}
\end{figure}

However one can follow also a different procedure which 
gives more physical insight and provides a further check of the result.
Let us consider the scattering of $N$ D0-branes out of the orientifold 
fixed-plane. 
If we separate out one of the D0-branes and calculate the force due to the other 
$N-1$ particles as depicted in fig.\ref{strings1} it is easy to see where the 
different 
powers of $N$ come from. Indeed a diagram like fig.\ref{strings2} is clarifying. 
Note 
that an extra power of $N$ is included to account for the fact that any of the 
$N$
particles can be singled out.
In other words, there are $N$ strings going from a D-particle to its own image 
and $N-1$ going from the D-particle to the image of some other D-particle.
It is clear then that one can perform the calculation for the case $N=2$
and then carefully separating the different contributions, multiply each one
by the corresponding factor of $N$. 
In this way the same result (\ref{eq:potential}) is 
obtained.

\section*{Appendix B}
\setcounter{section}{2}
\setcounter{equation}{0}

Here we list the expressions for the different diagrams.
The phase shift is given by
\be
\int d\tau_1 d\tau_2 g_i(\tau_1,\tau_2) ,
\ee
where $g_i(\tau_1,\tau_2)$ for different
diagrams are listed  below.
The  bosonic propagators are \cite{beckers,ggr}:
\begin{itemize}
\item{for a massive boson of mass $m^2=b^2 + (v \tau)^2+\delta$, }
\beqs
\p(\tau_1,\tau_2) &=&\int_{0}^{\infty} ds e^{-(b^2+\delta) s} \sqrt{\frac{v}{2 \pi 
\sinh{2 s v}}} \times \nonumber \\
&&\times \exp \left( \frac{v}{2 \sinh{2 s v}} \left( (\tau_1^2 +\tau_2^2)
\cosh{2 s v} - 2 \tau_1 \tau_2 \right) \right)
\eeqs
\item{for a massless boson}
\be
\p_0= -(\tau_1-\tau_2) \theta (\tau_1-\tau_2). 
\ee
\end{itemize}
For the fermionic diagrams we use that the fermionic propagator is related
to the bosonic one through 
\bea
\Delta_{ F} (\tau,\tau' | v\tau \gamma_9+b\gamma_8)&=&\left( 
\partial_{\tau} +v \tau \gamma_9 +b\gamma_8 \right) \Delta_{ B} \left
( \tau,\tau' | r^2-v\gamma_9 \right), \nonumber \\
\Delta_{ F} (\tau,\tau'|  0 )&=&  \partial_{\tau}
 \Delta_{ B} \left( \tau,\tau' | 0\right). 
\eea
The following notation is used for the different propagators
\be
\begin{array}{lcc}
 m^2=b^2 + (v \tau)^2+\delta,\,\delta \neq0 &: &\qquad \Delta_\delta  \\
 m^2=b^2 + (v \tau)^2 & : & \qquad \Delta  \\
 m=0 & : & \qquad \Delta_0 .
\end{array}
\ee

\begin{description}
\item{a) Diagram with a quartic vertex}
\bea
g_1 &=& 
N^3 \left\{ 2 \p (\pvv + \pmvv) +\frac{1}{4}\Delta_{2v}^2 
+\frac{1}{4}\Delta_{-2v}^2 
-\frac{1}{4}\pvv \pmvv    +7\p^2 \right\}  \nonumber \\
&&+ N^2 \left\{ 2 \p (\pvv + \pmvv) +\frac{1}{2}(\Delta_{2v}^2 +\Delta_{-2v}^2) 
-\frac{1}{2}\pvv \pmvv 
    \right\}  \nonumber \\
&& + N \left\{ \frac{1}{4}\Delta_{2v}^2 +\frac{1}{4}\Delta_{-2v}^2 
-\frac{1}{4}\pvv \pmvv 
 - 4 {\p}^2 \right\}.   
\eea

\item{ b0) Two bosonic cubic vertices without time derivatives}
\bea
g_2&=& \frac{N^3}{8} \bigg\{ b^2 \p_0  \left[ 2 \p (\pvv + \pmvv) +4 
(\Delta_{2v}^2 
+\Delta_{-2v}^2)
 +42 {\p}^2  \right] \nonumber \\
&&  +v^2 \tau_1 \tau_2 \p_0 \left[  ( 8 \p (\pvv + \pmvv) +\Delta_{2v}^2 
  +\Delta_{-2v}^2
 +4 \pvv \pmvv +32 {\p}^2 \right] \bigg\}
 \nonumber \\
&&+ \frac{N^2}{8} \bigg\{ b^2 \p_0  \left[ 4 \p (\pvv + \pmvv) +8 (\Delta_{2v}^2 
+\Delta_{-2v}^2)
 +4 {\p}^2  \right] \nonumber \\
&&  +v^2 \tau_1 \tau_2 \p_0 \left[  ( 8 \p (\pvv + \pmvv) +2 (\Delta_{2v}^2 
  +\Delta_{-2v}^2)
 +8 \pvv \pmvv \right] \bigg\}+
 \nonumber \\
&&+ \frac{N}{8} \bigg\{ b^2 \p_0 \left[ -2 \p (\pvv + \pmvv) -4 (\Delta_{2v}^2 
+\Delta_{-2v}^2)
 -26 {\p}^2 \right] \nonumber \\
&&  +v^2 \tau_1 \tau_2 \p_0 \left[ -32 \p^2 - (\Delta_{2v}^2 
  +\Delta_{-2v}^2)
 -4 \pvv \pmvv \right] \bigg\} .
\eea

\item{ b1) Two bosonic cubic vertices with one time derivative}
 
\bea
g_3&=& \frac{N^3}{4} \bigg\{ v \tau_1  \bigg\{ - (\dii \p_0 )(\Delta_{2v}^2 - 
\Delta_{-2v}^2)
+  \p_0 (\pvv + \pmvv)\dii (\pvv - \pmvv ) \nonumber \ \\
&& - \p_0  (\pvv \dii \pmvv - \pmvv \dii \pvv)  \nonumber \\
&& -8 (\pvv - \pmvv) (\p_0 \dii \p - \p \dii \p_0) \bigg\} \bigg\}
\nonumber \\
&&+ \frac{N^2}{4} \bigg\{ v \tau_1  \bigg\{ -2 (\dii \p_0 )(\Delta_{2v}^2 - 
\Delta_{-2v}^2)
+ 2 \p_0 (\pvv + \pmvv)\dii (\pvv - \pmvv ) \nonumber \ \\
&& -2 \p_0  (\pvv \dii \pmvv - \pmvv \dii \pvv)  \nonumber \\
&& -8 (\pvv - \pmvv) (\p_0 \dii \p - \p \dii \p_0) \bigg\} \bigg\}
\nonumber \\
&& + \frac{N}{4} \bigg\{v \tau_1  \bigg\{ - \dii \p_0 (\Delta_{2v}^2 - 
\Delta_{-2v}^2)
+ \p_0 (\pvv + \pmvv)\dii (\pvv - \pmvv ) \nonumber \ \\
&& - \p_0  (\pvv \dii \pmvv - \pmvv \dii \pvv)   \bigg\} \bigg\} .
\eea

\item{ c2) Two bosonic cubic vertices with two time derivatives}
\bea
g_4&=& -\frac{N^3}{8} \bigg\{ 16 \p_0 (\di \dii \p) \p -16 \p_0 (\di \p) (\dii 
\p) 
-2 \p_0  (\di \pvv) (\dii\pmvv)  \nonumber \\
&& +  \p_0 (\pvv + \pmvv)  \di \dii (\pvv + \pmvv)
+8 \p (\pvv + \pmvv) \di \dii \p_0 \nonumber \\
&& +8 \p_0 (\pvv + \pmvv) \di \dii \p
-16 \di \p_0 \dii \p (\pvv + \pmvv) \nonumber \\
&& -2 (\di \p_0)  (\pvv \dii \pvv + \pmvv \dii \pmvv) 
 + (\Delta_{2v}^2 +\Delta_{-2v}^2) \di \dii \p_0 \bigg\} \nonumber \\
&&- \frac{N^2}{8} \bigg\{  2 \p_0 (\pvv + \pmvv)  \di \dii (\pvv + \pmvv) 
\nonumber \\
&& -4 \p_0  (\di \pvv) (\dii\pmvv) 
+8 \p (\pvv + \pmvv) \di \dii \p_0 \nonumber \\
&& +8 \p_0 (\pvv + \pmvv) \di \dii \p
-16 \di \p_0 \dii \p (\pvv + \pmvv) \nonumber \\
&& -4 (\di \p_0)  (\pvv \dii \pvv + \pmvv \dii \pmvv) 
 +2 (\Delta_{2v}^2 +\Delta_{-2v}^2) \di \dii \p_0 \bigg\}\nonumber \\ 
&& + \frac{N}{8} \bigg\{  -\p_0 (\pvv + \pmvv)  \di \dii (\pvv + \pmvv) 
\nonumber \\
&& +2 \p_0  (\di \pvv) (\dii\pmvv) 
-16 \p \p_0 \di \dii \p \nonumber \\
&& + 16 \di \p \dii \p \p_0
+16 \di \p_0 \dii \p (\pvv + \pmvv) \nonumber \\
&& +2 (\di \p_0)  (\pvv \dii \pvv + \pmvv \dii \pmvv) 
 - (\Delta_{2v}^2 +\Delta_{-2v}^2) \di \dii \p_0 \bigg\} . \nonumber \\
\eea

\item{ c0) Two ghost vertices without time derivatives}
\bea
g_5&=& - (b^2 + v \tau_1 \tau_2) \p_0 \p \p \frac{N^3}{4} \nonumber \\
&& - (b^2 + v \tau_1 \tau_2) \p_0 \p \p \frac{N^2}{2} \nonumber \\
&& - (b^2 + v \tau_1 \tau_2) \p_0 \p \p \frac{N}{4} .
\eea

\item{ c1) Two ghost vertices with one time derivative}
\bea
g_6&=& v \tau_1 (\dii \p_0) \p (\pvv- \pmvv) \frac{N^3}{4} \nonumber \\
&& + v \tau_1 (\dii \p_0) \p (\pvv- \pmvv) \frac{N^2}{2} \nonumber \\
&& + v \tau_1 (\dii \p_0) \p (\pvv- \pmvv) \frac{N}{4} .
\eea

\item{ c2) Two ghost vertices with two time derivatives}
\bea
g_7&=& \left[  (\di \p_0) (\dii \p) (\pvv + \pmvv) + \p_0 (\di \p) (\dii \p)
\right] \frac{N^3}{4}\nonumber \\
&& +\left[  (\di \p_0) (\dii \p) (\pvv + \pmvv) + \p_0 (\di \p) (\dii \p)
\right] \frac{N^2}{2}\nonumber \\
&& + \left[  (\di \p_0) (\dii \p) (\pvv + \pmvv) + \p_0 (\di \p) (\dii \p)
\right] \frac{N}{4} .
\eea

 \item{ d) Two fermionic vertices }
\bea
g_8&=& \frac{N^3}{8}    \p_0 \bigg\{ - 4  \left[ ( \di \pv )( \dii \pv) - (\di 
\pmv) (\dii 
\pmv )
\right] - 64 \di \pv \dii \pmv \nonumber \\
&&+8 v \tau_1 (\pv \dii \pv - \pmv \dii \pmv)
+64  v \tau_1 (\pv \dii \pmv - \pv \dii \pmv) \nonumber \\
&& +v^2 \tau_1 \tau_2 \left( - 4(\pv^2 +\pmv^2) +64 \pv \pmv \right)
+b^2 \left( +24 (\pv^2 +\pmv^2) +8 \pv \pmv \right)
 \nonumber \\
&& + 64 ( \di \p_0 ) \p \left[ \dii (\pv + \pmv) +v \tau_2 (\pv - \pmv) \right]
\bigg\}   \nonumber \\ 
&& + 0 \cdot N^2 \nonumber \\
&&+\frac{N}{8}  \p_0 \bigg\{ - 2  \left[ ( \di \pv )( \dii \pv) - (\di \pmv) 
(\dii \pmv )
\right] - 32 \di \pv \dii \pmv \nonumber \\
&&+4 v \tau_1 (\pv \dii \pv - \pmv \dii \pmv)
+32  v \tau_1 (\pv \dii \pmv - \pv \dii \pmv) \nonumber \\
&& +v^2 \tau_1 \tau_2 \left( - 2(\pv^2 +\pmv^2) +32 \pv \pmv \right)
+b^2 \left( +8 (\pv^2 +\pmv^2) +8 \pv \pmv \right)
 \nonumber \\
&& + 32 ( \di \p_0 ) \p \left[ \dii (\pv + \pmv) +v \tau_2 (\pv - \pmv) \right]
\bigg\} .
\eea

\end{description}

\end{document}